\documentclass[twoside]{ilcws07}
\usepackage[latin1]{inputenc}
\usepackage[dvips]{graphicx,epsfig,color}
\usepackage{wrapfig,rotating}
\usepackage{amssymb,amsmath,array}

\newcommand{\Wev} {$W \rightarrow e \nu$}
\newcommand{\Wuv} {$W \rightarrow \mu \nu$}
\newcommand{\Zee} {$Z \rightarrow e^+ e^-$}
\newcommand{\Zuu} {$Z \rightarrow \mu^+ \mu^-$}

\newcommand{\C} {\v{C}erenkov }
\newcommand{\dream} {{\sc dream}}

\begin{document}
\title{Muon Identification without Iron} 
\author{John Hauptman
\thanks{This work has been supported by the US Department of Energy
through DE-FG02-91ER40634 and LCRD Project 6.18.}
\vspace{.3cm}\\
Iowa State University, Department of Physics and Astronomy \\
Ames, IA  50011  USA
}

\maketitle

\begin{abstract}
Muons can be identified with high efficiency and purity and reconstructed 
with high precision is a detector with a dual readout calorimeter and a 
dual solenoid to return the flux without iron.  We shown CERN test beam
data for the calorimeter and calculations for the magnetic fields and the
track reconstruction.  For isolated tracks, the rejection of pions against muons
ranges from $10^3$ at 20 GeV/c to $10^5$ at 300 GeV/c.
\end{abstract}

\section{Introduction}

\begin{wrapfigure}{c}{0.6\columnwidth}
\centerline{\includegraphics[width=0.55\columnwidth]{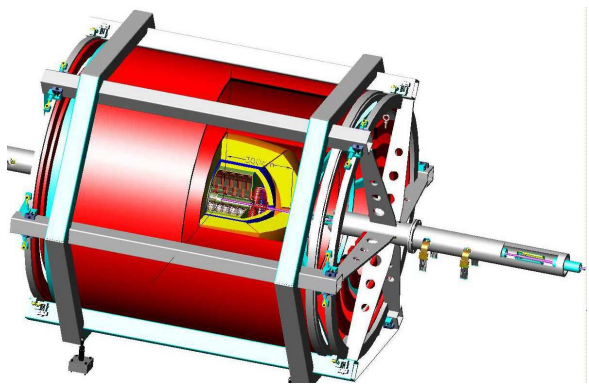}}
\caption{4h Concept detector showing the dual solenoids.  The annulus between the solenoids is
 filled with cluster counting wires inside precision tubes. }\label{Fig:4th}
\end{wrapfigure}

Big detectors at high energy colliders require the detection of electrons ($e$)
and muons ($\mu$)
with high efficiency, high purity and high precision for the reconstruction of the decays
\Wev, \Wuv, \Zee, \Zuu,  for the identification of $\tau$ lepton decays to $e$ and
$\mu$, for searches for lepton number violation, for the positive tagging of events with
missing neutrinos, and for the isolation of event samples
with supposed decays of supersymmetric or other massive states decaying partly to
leptons.   The muon system of
the 4th Concept,  Fig. \ref{Fig:4th},  achieves almost absolute muon identification for isolated tracks.  

\section{Separation of muons ($\mu^{\pm}$) from charged pions ($\pi^{\pm}$)}

We achieve excellent $\mu-\pi^{\pm}$ separation using three independent
measurements: (a)  energy balance
from the tracker through the calorimeter into the muon spectrometer, (b) a
unique separation of the radiative component from the ionization component
in the dual-readout calorimeter; and, (c) a measurement of the neutron content
in the dual-readout fiber calorimeter.  

\subsection{Separation energy and momentum balance}

The central tracking system has a resolution of about 
$\sigma_p/p^2 \sim k_1 $  ($k_1 \sim 3 \times 10^{-5}$ (GeV/c)$^{-1}$)
and the muons spectrometer in the annulus
between the solenoids with a B=1.5 T field has resolution of about 
$\sigma_p/p^2 \sim k_2 $   ($k_2 \sim 5 \times 10^{-4}$ (GeV/c)$^{-1}$).    A muon of momentum $p$ 
 which radiates energy $E$ in the
volume of the calorimeter that is measured with a resolution of $\sigma_E/E \sim 0.20/\sqrt{E}$ 
 will have a momentum-energy balance constraint of 
 $[k_1 p^2 \oplus 0.2\sqrt{E} \oplus k_2 (p-E)^2]/p$ which yields a rejection of about 30 for a 100 GeV
 muon radiating 20 GeV.
 
\begin{figure}[hbtp]
  \vspace{9pt}

  \centerline{\hbox{ \hspace{0.0in} 
    \epsfxsize=2.5in
    \epsffile{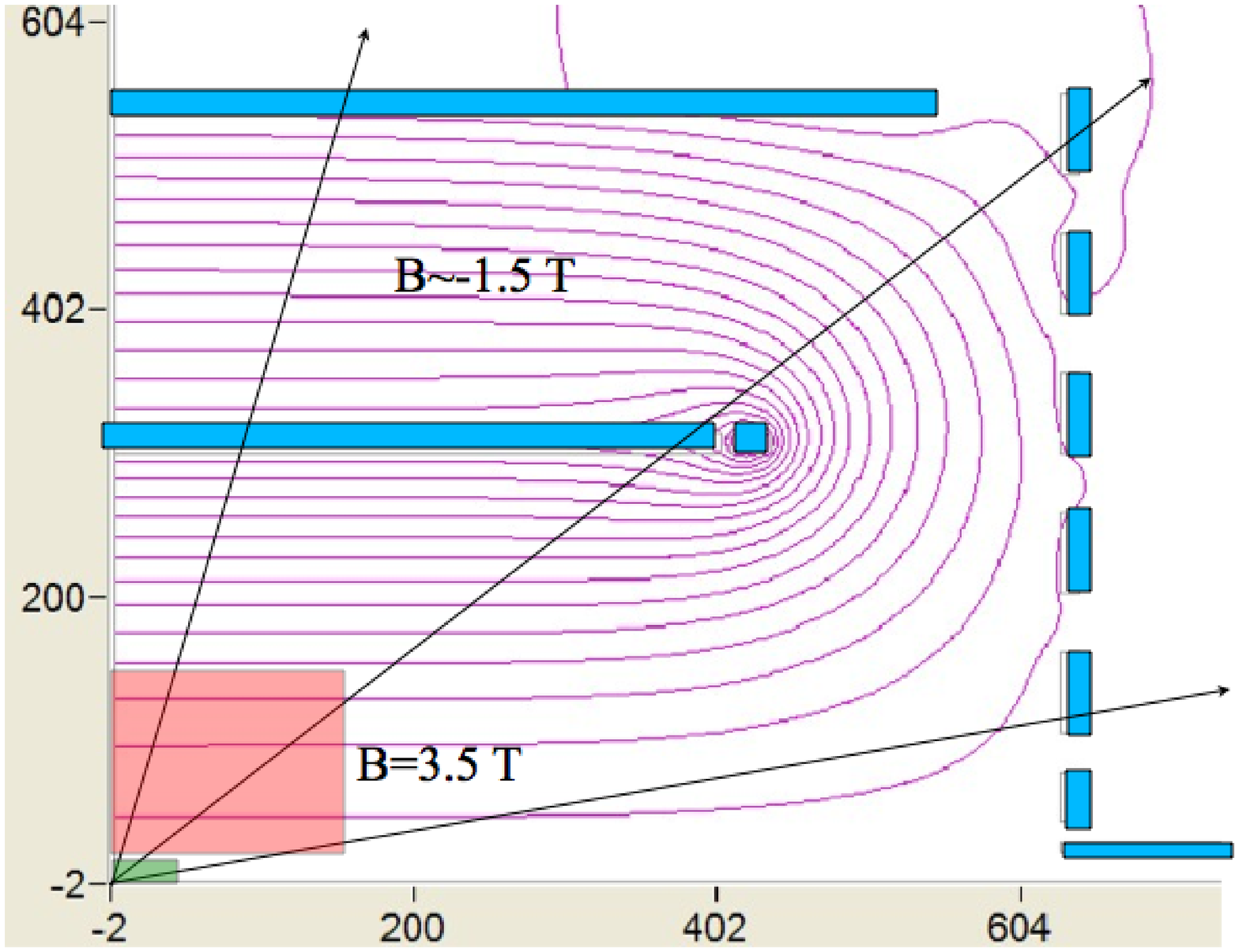}
    \hspace{0.15in}
    \epsfxsize=2.5in
    \epsffile{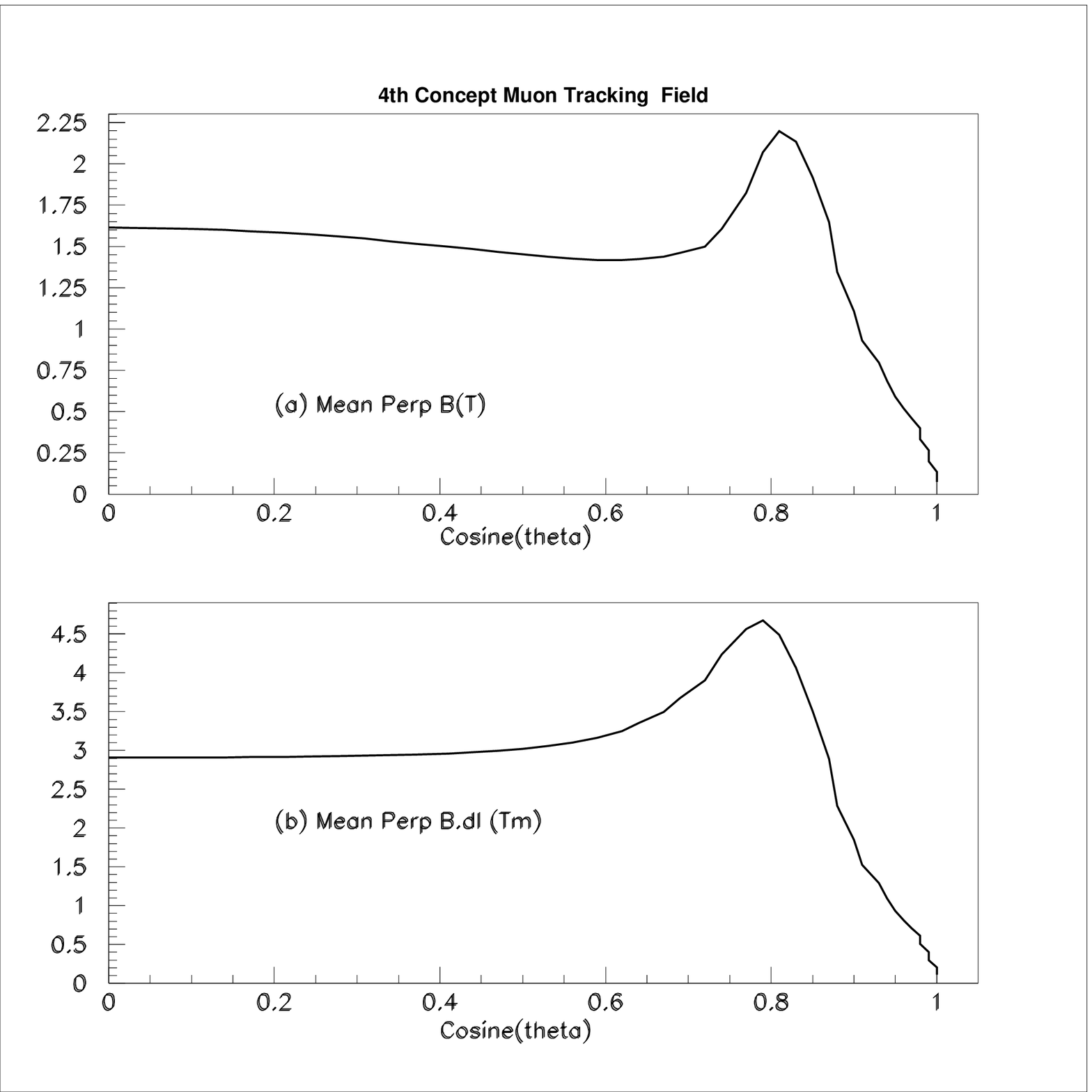}
    }
  }

  \vspace{9pt}
  \hbox{\hspace{1.35in} (a) \hspace{2.10in} (b)} 
  \vspace{9pt}
  \caption{(a) The field map of the dual solenoids showing the wall of coils and the almost completely confined field; (b) the mean bending field (1.5 T) along muon trajectories from the origin and the integral of the mean bending field (3 Tm).}
  \label{fig:Bmu}

\end{figure}

\subsection{Separation by dual-readout}

\begin{wrapfigure}{r}{0.7\columnwidth}
\centerline{\includegraphics[width=0.65\columnwidth]{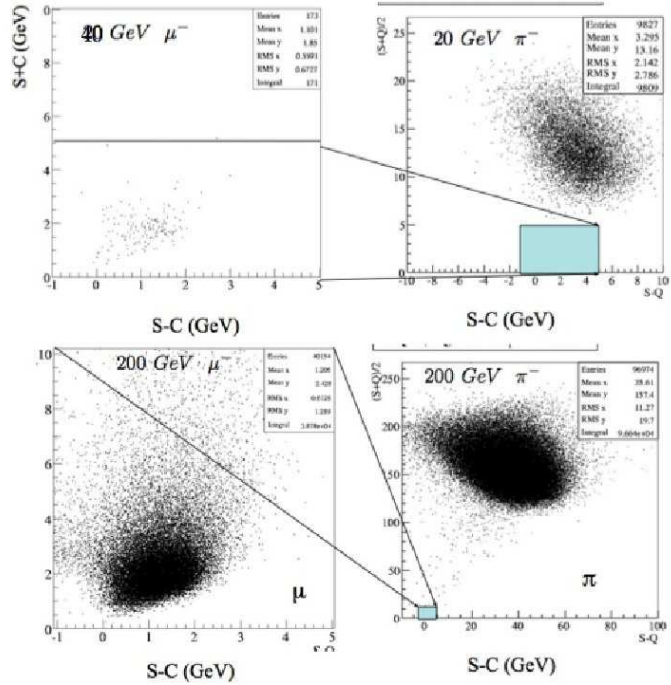}}
\caption{Dual-readout separation of $\mu-\pi^{\pm}$.}\label{Fig:mu-pi}
\end{wrapfigure}

A non-radiating muon penetrating the mass of a fiber dual readout calorimeter will leave a signal in the scintillating fibers equivalent to the $dE/dx$ of the muon, which in \dream ~ is about 1.1 GeV.  There will be no \C signal since the \C angle is larger than the capture cone angle of the fiber.  A radiating muon will add equal signals to both the scintillating and \C fibers and, therefore, the difference of the scintillating $(S)$ and \C $(C)$ signals is
\begin{displaymath}
  S-C \approx ~  \overline{dE/dx} \approx 1.1 ~ GeV
\end{displaymath}
independent of the degree of radiation.  The distributions of $(S-C)$
 {\it vs.} $(S+C)/2$ for 20 GeV\footnote{There were no $\mu^-$ left in the H4 beam at 20 GeV/c, so these data are from 40 GeV/c $\mu^-$.}   
  and 200 GeV $\pi^-$ and $\mu^-$ are shown in Fig.
 \ref{Fig:mu-pi} in which for an isolated track the $\pi^{\pm}$ 
 rejection against $\mu$ is about $10^3$ at 20 GeV and $10^4$ at 200 GeV.  The distribution of $(S-C)$ 
 as a mean that is very nearly 1.1 GeV as expected, and the radiative events are evident at larger 
 $(S+C)/2$.

\medskip

\subsection {Separation by neutron content measurement}

The \dream ~  collaboration has succeeded in the measurement of neutron content in hadronic
showers event-by-event in the \dream ~  module by summing the scintillating
channels of the module in three radial rings and digitizing the PMT output  at 1.25 GHz.  These data,
now being analyzed, show clearly the long-time neutron component in hadron showers that
is absent in electromagnetic showers (and also absent in the \C fibers of the \dream ~ module
for both $e$ and $\pi$).

We expect to estimate a neutron fraction, $f_n$, each event the same way we estimate
$f_{EM}$  each event, and be able to reject localized hadronic activity in the calorimeter with
 factors of 10-50. 

\section{Summary}

Any simple product of these three rejection factors, or any estimate of the muon efficiency and purity,
gives an optimistic result that will clearly be limited by tracking efficiencies, overlapping shower debris 
in the calorimeter, or track confusion either in the main tracker or the muon spectrometer before these
beam test numbers are reached.  Nevertheless, we expect excellent muon identification.  



\begin{footnotesize}

\end{footnotesize}

\end{document}